\documentclass[preprint,number]{elsarticle}

\usepackage{graphicx}
\usepackage{color}
\usepackage[dvipsnames]{xcolor}
\usepackage{xfrac}
\usepackage{epsfig}
\usepackage{hyperref}
\usepackage{multicol}
\usepackage[all]{hypcap}
\usepackage[letterpaper,bindingoffset=0.0in,left=1in,right=1in,top=1in,bottom=1in,footskip=.25in]{geometry}
\usepackage{xspace}
\usepackage{booktabs}
\usepackage{amsmath}
\usepackage{multirow}
\usepackage{subcaption}
\usepackage{csquotes}

\makeatletter
\def\ps@pprintTitle{%
 \let\@oddhead\@empty
 \let\@evenhead\@empty
 \def\@oddfoot{}%
 \let\@evenfoot\@oddfoot}
\makeatother

\usepackage{xpatch}
\xpatchcmd{\MaketitleBox}{\hrule}{}{}{}
\xpatchcmd{\MaketitleBox}{\hrule}{}{}{}
\xpatchcmd{\pprintMaketitle}{\hrule}{}{}{}
\xpatchcmd{\pprintMaketitle}{\hrule}{}{}{}

\usepackage{etoolbox}
\patchcmd{\abstract}{Abstract}{\vspace{-\baselineskip}}{}{}

\newcommand{\cenns}{CENNS-10\xspace}
\newcommand{\cevns}{\protect{CEvNS}\xspace}
\newcommand{\etal}{\emph{et al.}\xspace}
\newcommand{\us}{\protect{\ensuremath{\mu\text{s}}}\xspace}

\newenvironment{keeptogether}{\par\noindent\begin{minipage}{\linewidth}}{\end{minipage}\par}

\usepackage{listings}
\definecolor{mygreen}{rgb}{0,0.6,0}
\definecolor{mygray}{rgb}{0.5,0.5,0.5}
\definecolor{mymauve}{rgb}{0.58,0,0.82}
\newcommand\YAMLcolonstyle{\color{red}\mdseries}
\newcommand\YAMLkeystyle{\color{black}\bfseries}
\newcommand\YAMLvaluestyle{\color{blue}\mdseries}

\makeatletter

\newcommand\language@yaml{yaml}

\expandafter\expandafter\expandafter\lstdefinelanguage
\expandafter{\language@yaml}
{
  keywords={true,false,null,y,n},
  keywordstyle=\color{darkgray}\bfseries,
  basicstyle=\YAMLkeystyle,                                 
  sensitive=false,
  comment=[l]{\#},
  morecomment=[s]{/*}{*/},
  commentstyle=\color{purple}\ttfamily,
  stringstyle=\YAMLvaluestyle\ttfamily,
  moredelim=[l][\color{orange}]{\&},
  moredelim=[l][\color{magenta}]{*},
  moredelim=**[il][\YAMLcolonstyle{:}\YAMLvaluestyle]{:},   
  morestring=[b]',
  morestring=[b]",
  literate =    {---}{{\ProcessThreeDashes}}3
                {>}{{\textcolor{red}\textgreater}}1     
                {|}{{\textcolor{red}\textbar}}1 
                {\ -\ }{{\mdseries\ -\ }}3,
}

\lst@AddToHook{EveryLine}{\ifx\lst@language\language@yaml\YAMLkeystyle\fi}
\makeatother

\newcommand\ProcessThreeDashes{\llap{\color{cyan}\mdseries-{-}-}}

\begin{document}

\title{COHERENT Collaboration data release from the first detection of coherent elastic neutrino-nucleus scattering on argon}

\address[itep]{Institute for Theoretical and Experimental Physics named by A.I. Alikhanov of National Research Centre ``Kurchatov Institute'', Moscow, 117218, Russian Federation}
\address[mephi]{National Research Nuclear University MEPhI (Moscow Engineering Physics Institute), Moscow, 115409, Russian Federation}
\address[iu]{Department of Physics, Indiana University, Bloomington, IN, 47405, USA}
\address[duke]{Department of Physics, Duke University, Durham, NC 27708, USA}
\address[tunl]{Triangle Universities Nuclear Laboratory, Durham, North Carolina, 27708, USA}
\address[utk]{Department of Physics and Astronomy, University of Tennessee, Knoxville, TN 37996, USA}
\address[ornl]{Oak Ridge National Laboratory, Oak Ridge, TN 37831, USA}
\address[sandia]{Sandia National Laboratories, Livermore, CA 94550, USA}
\address[uwcenpa]{Department of Physics and Center for Experimental Nuclear Physics and Astrophysics,\\ University of Washington, Seattle, WA 98195, USA}
\address[usd]{Physics Department, University of South Dakota, Vermillion, SD 57069, USA}
\address[nmsu]{Department of Physics, New Mexico State University, Las Cruces, NM 88003, USA}
\address[lanl]{Los Alamos National Laboratory, Los Alamos, NM, USA, 87545, USA}
\address[ncsu]{Physics Department, North Carolina State University, Raleigh, NC 27695, USA}
\address[vt]{Center for Neutrino Physics, Virginia Tech, Blacksburg, VA 24061, USA}
\address[nccu]{Department of Mathematics and Physics, North Carolina Central University, Durham, NC, 27707, USA}
\address[cmu]{Carnegie Mellon University, Pittsburgh, PA 15213, USA}
\address[ufl]{Department of Physics, University of Florida, Gainesville, FL 32611, USA}
\address[efi]{Enrico Fermi Institute, University of Chicago, Chicago, IL 60637, USA}
\address[kicp]{Kavli Institute for Cosmological Physics, University of Chicago, Chicago, IL 60637, USA}
\address[mipt]{Moscow Institute of Physics and Technology, Dolgoprudny, Moscow Region 141700, Russian Federation}
\address[laurentian]{Department of Physics, Laurentian University, Sudbury, Ontario P3E 2C6, Canada}
\address[tufts]{Department of Physics and Astronomy, Tufts University, Medford, MA 02155, USA}

\address[kaist]{Department of Physics at Korea Advanced Institute of Science and Technology (KAIST), Daejeon, 34141, Republic of Korea}
\address[ibs]{Center for Axion and Precision Physics Research (CAPP) at Institute for Basic Science (IBS), Daejeon, 34141, Republic of Korea}

\author[itep,mephi]{D.~Akimov}
\author[iu]{J.B.~Albert}
\author[duke,tunl]{P.~An}
\author[duke,tunl]{C.~Awe}
\author[duke,tunl]{P.S.~Barbeau}
\author[utk]{B.~Becker}
\author[itep,mephi]{V.~Belov}
\author[ornl]{M.A.~Blackston}
\author[utk]{L.~Blokland}
\author[mephi]{A.~Bolozdynya}
\author[sandia]{B.~Cabrera-Palmer}
\author[uwcenpa]{N.~Chen}
\author[usd]{D.~Chernyak}
\author[duke]{E.~Conley}
\author[nmsu,lanl]{R.L.~Cooper}
\author[utk]{J.~Daughhetee}
\author[iu]{M.~del~Valle~Coello}
\author[uwcenpa]{J.A.~Detwiler}
\author[uwcenpa]{M.R.~Durand}
\author[utk,ornl]{Y.~Efremenko}
\author[lanl]{S.R.~Elliott}
\author[ornl]{L.~Fabris}
\author[ornl]{M.~Febbraro}
\author[iu]{W.~Fox}
\author[utk,ornl]{A.~Galindo-Uribarri}
\author[tunl,ornl,ncsu]{M.P.~Green}
\author[uwcenpa]{K.S.~Hansen}
\author[ornl]{M.R.~Heath}
\author[duke,tunl]{S.~Hedges}
\author[iu]{M.~Hughes}
\author[duke,tunl]{T.~Johnson}
\author[nmsu]{M.~Kaemingk}
\author[iu]{L.J.~Kaufman \fnref{kauf}}
\fntext[kauf]{Presently at SLAC National Accelerator Laboratory, Menlo Park, CA 94205, USA}
\author[mephi]{A.~Khromov}
\author[itep,mephi]{A.~Konovalov}
\author[itep,mephi]{E.~Kozlova}
\author[mephi]{A.~Kumpan}
\author[duke,tunl]{L.~Li}
\author[uwcenpa]{J.T.~Librande}
\author[vt]{J.M.~Link}
\author[usd]{J.~Liu}
\author[tunl,ncsu]{K.~Mann}
\author[tunl,nccu]{D.M.~Markoff}
\author[uwcenpa]{O.~McGoldrick}
\author[nmsu]{H.~Moreno}
\author[ornl]{P.E.~Mueller}
\author[ornl]{J.~Newby}
\author[cmu]{D.S.~Parno}
\author[ornl]{S.~Penttila}
\author[duke]{D.~Pershey}
\author[ornl]{D.~Radford}
\author[cmu]{R.~Rapp}
\author[ufl]{H.~Ray}
\author[duke]{J.~Raybern}
\author[itep,mephi]{O.~Razuvaeva}
\author[sandia]{D.~Reyna}
\author[efi,kicp]{G.C.~Rich}
\author[itep,mephi]{D.~Rudik}
\author[duke,tunl]{J.~Runge}
\author[iu]{D.J.~Salvat}
\author[duke]{K.~Scholberg}
\author[mephi]{A.~Shakirov}
\author[itep,mephi,mipt]{G.~Simakov}
\author[duke]{G.~Sinev}
\author[iu]{W.M.~Snow}
\author[mephi]{V.~Sosnovtsev}
\author[iu]{B.~Suh}
\author[iu]{R.~Tayloe}
\author[vt]{K.~Tellez-Giron-Flores}
\author[iu,lanl]{R.T.~Thornton}
\author[iu]{I.~Tolstukhin \fnref{tols}}
\fntext[tols]{Presently at Argonne National Laboratory, Argonne, IL 60439, USA}
\author[iu]{J.~Vanderwerp}
\author[ornl]{R.L.~Varner}
\author[laurentian]{C.J.~Virtue}
\author[iu]{G.~Visser}
\author[uwcenpa]{C.~Wiseman}
\author[tufts]{T.~Wongjirad}
\author[tufts]{J.~Yang}
\author[cmu]{Y.-R.~Yen}
\author[kaist,ibs]{J.~Yoo}
\author[ornl]{C.-H.~Yu}
\author[iu]{J.~Zettlemoyer}


\begin{abstract}
The enclosed data release includes the information to analyze the COHERENT data published in Ref.~\cite{Akimov:2020}. The data, the \cevns signal, and the associated backgrounds are shared in a binned text file format along with associated uncertainties. The binning of the data in the text file is identical to that in Ref.~\cite{Akimov:2020}. This document provides information on the enclosed data release and guidance on the use of the data. 
\end{abstract}

\maketitle

\section{Overview of the release}
\subsection{Accessing the release}
This data release follows ``Analysis A'' of Ref.~\cite{Akimov:2020} with the information included within the release and in this document. As ``Analysis B'' gives consistent results as reported in Ref.~\cite{Akimov:2020}, it is not reported as part of this release. The data release, this accompanying document, and code examples provided with the release are available in two locations: at \url{http://coherent.ornl.gov/data} and also on Zenodo (DOI: \href{http://dx.doi.org/10.5281/zenodo.3903810}{{\tt 10.5281/zenodo.3903810}}).
See Sec.~\ref{sec:prescription} for comments on how to use the released data. See Sec.~\ref{sec:citing} for how to cite this data. Please direct questions about the material provided within this release to \texttt{jzettle@fnal.gov} (J. Zettlemoyer) and/or \texttt{rtayloe@indiana.edu} (R. Tayloe).

\subsection{Materials Provided}
There are two main methods of distributing the relevant information in this data release. The provided data and signal/background probability distribution functions (PDFs) are included as 3D binned arrays in text file formats. They are identically binned in the 3D space in energy, $F_{90}$, and time to trigger ($t_{\mathrm{trig}}$) as in Ref.~\cite{Akimov:2020}. The binning is given in Table~\ref{tab:binning}. Values such as the neutrino flux with associated uncertainties are provided in a YAML file format as in the previous collaboration data release~\cite{Akimov:2018vzs} accompanying the first observation of \cevns in Ref.~\cite{Akimov:2017ade}. The YAML format and the parameter values included are described in Sec.~\ref{sec:yaml}. All files described here are located in the \texttt{Data} directory of this release.

For analyses that do not require the entire 3-dimensional information, 1-dimensional projections in the form of binned text files that include all the information to recreate Fig.~4 of Ref.~\cite{Akimov:2020} are also included as part of this data release with the format of the lines described at the top of the file. These files are located in the \texttt{Data/OneDProjections} directory of the release and are labeled \texttt{energydata1d.txt} for energy,\texttt{f90data1d.txt} for $F_{90}$, and \texttt{timingdata1d.txt} for $t_{\mathrm{trig}}$.  

\begin{table}
\centering
\begin{tabular}[ht!]{l l r r r}
    \toprule
	Dimension & Unit & Range & Bin Width & Total Bins \\
	\midrule
	Energy & keVee & 0 -- 120 & 10 & 12\\
	$F_{90}$ & $F_{90}$ & 0.5 -- 0.9 & 0.05 & 8\\
	$t_{\mathrm{trig}}$ & \us & -0.1 -- 4.9 & 0.5 & 10\\
	\bottomrule
	\end{tabular}
	\caption{}
    \label{tab:binning}
\end{table}

\section{Included within release} \label{sec:included}
A description of the major contents of the release is below.

\subsection{SNS Data}
For the SNS data, both ``on-beam'' (SNS beam data, \texttt{datanobkgsub.txt}) and ``off-beam'' (measured steady-state backgrounds, \texttt{bkgpdf.txt}) triggered data are included separately. Both are provided as a tab-separated-value text file with in the form of: bin center in energy[keVee], bin center in $F_{90}$[$F_{90}$], bin center in $t_{\mathrm{trig}}$[\us], number of events/bin/6.12 GWhr. The number of total bins and bin widths in each dimension is the same as for Analysis A in Ref.~\cite{Akimov:2020} and described in Tab.~\ref{tab:binning}. 

\subsection{\cevns Signal PDF}
The \cevns signal PDF used in Analysis A of Ref.~\cite{Akimov:2020} is included as part of this data release. The provided text file (\texttt{cevnspdf.txt}) is normalized to the initial central value (CV) SM prediction from Analysis A of the COHERENT data. This information is provided in the same way as the SNS data. The normalizations for both the best-fit result and the SM prediction are included in the YAML file describing single-value parameters. The \cevns signal was allowed to float during the likelihood fit in the analysis of Ref.~\cite{Akimov:2020} but an uncertainty is included representing the cross-section systematic errors as in Tab.~2 of Ref.~\cite{Akimov:2020}.

This release also includes information on the measured energy resolution and the SNS timing parameters. The SNS protons-on-target trace is well-approximated by a Gaussian distribution. The YAML file includes the mean and width of the distribution and the uncertainties on those values with respect to $t_{\mathrm{trig}}$. The $F_{90}$ distribution can be examined by looking at the 2-D projection in $F_{90}$ and energy within the binned PDF \texttt{cevnspdf.txt}. 

\subsection{Beam-related Neutrons (BRN)}
The prompt BRN PDF from Analysis A of Ref.~\cite{Akimov:2020} is included within the release. As they are treated as separate components during the likelihood analysis in Ref.~\cite{Akimov:2020}, the delayed BRN PDF is included separately. As for the \cevns PDF, the prompt and delayed BRN PDFs is normalized within the text files (\texttt{brnpdf.txt} for prompt BRN, \texttt{delbrnpdf.txt} for delayed BRN) to the initial central value (CV) prediction from Analysis A. An uncertainty value located in the YAML file represents the width of a Gaussian constraint The normalization for the best-fit result is also included within the YAML file and can be renormalized if needed.

\subsection{Steady-state Backgrounds}
The steady-state background (\texttt{bkgpdf.txt}) PDF is the measured off-beam data. The separate off-beam trigger computes the steady-state contribution \textit{in situ}. The energy and $F_{90}$ components of the PDF come directly from the measured off-beam data. The time component is included as a constant over the considered time range. The steady-state PDF is normalized within the text file (\texttt{bkgpdf.txt}) to the CV prediction from Analysis A. An uncertainty value which represents the width of a Gaussian constraint is included in the YAML file. The normalization for the best-fit result is also included within the YAML file and can be renormalized if needed.

Note that the steady-state background is originally oversampled from a 5x larger window with respect to $t_{\mathrm{trig}}$ than the data to reduce the statistical error on the background. This has been taken into account in the provided PDF and the normalization includes this information. However, when a subtraction of the steady-state background is performed on the data within an analysis, the consequence of the oversampling must be applied. To do so, the error on each bin is not $\sqrt{N}$, but $\frac{\sqrt{5N}}{5}$. 

\subsection{Summary up to now}
A summary of the information described in this section is given in Tab.~\ref{tab:summaryinfo}. 

\begin{table}
\centering
\begin{tabular}[ht!]{l l r r r}
    \toprule
	Distribution & File & CV Prediction & Best-fit result \\
	\toprule
	On-beam Data & \texttt{datanobkgsub.txt} & \multicolumn{2}{c}{3752} \\
	\midrule
	\cevns & \texttt{cevnspdf.txt} & $128\pm17$ & $159\pm43$ \\
	BRN, prompt & \texttt{brnpdf.txt} & $497\pm160$ & $553\pm34$ \\
	BRN, delayed & \texttt{delbrnpdf.txt} & $33\pm33$ & $10\pm11$ \\
	Steady-state background & \texttt{bkgpdf.txt} & $3152\pm25$ & $3131\pm23$ \\
	\bottomrule
	\end{tabular}
	\caption{Summary of information provided in the text files}
    \label{tab:summaryinfo}
\end{table}

\subsection{Detector Efficiency}
The detector efficiency after cuts is included as a separate text file (\texttt{CENNS10AnlAEfficiency.txt}) in energy space with lines: bin center in keVee, bin center in keVnr, efficiency. This includes effects of cuts in $F_{90}$ space corresponding to those used in Analysis A of Ref.~\cite{Akimov:2020}. Use a flat efficiency value corresponding to the value in the last bin for any reconstructed energies larger than those given in the text file. 

\subsection{Energy Resolution}
The energy resolution is well described by $\frac{\sigma_E}{E}=\frac{\mathrm{a}}{\sqrt{E(keVee)}}$ and included as part of the data release inside the YAML file. The value of the parameter a is determined from the $^{83m}$Kr calibration data and confirmed using the \cenns simulation.

\subsection{Single-Value/Functional Parameters} \label{sec:yaml}
A YAML file (\texttt{LArParametersAnlA.yaml}) represents parameters represented with a single-value or a functional form included within this release. An entry in the YAML file includes the parameter of function values, uncertainties, and a comment describing the parameter of function. In the case of a function the functional form is provided in the comment. The parameters included in the YAML file are:
\begin{itemize}
    \item Beam exposure
    \item Distance to SNS target
    \item Detector mass
    \item Quenching Factor (QF)
    \item $\nu$/proton
    \item Best-fit normalizations with uncertainties for \cevns, prompt and delayed BRN, and steady-state background
    \item Initial CV prediction normalizations with prior uncertainties for \cevns, prompt and delayed BRN, and steady-state background
    \item Detector efficiency
    \item Energy Resolution
    \item SNS protons-on-target timing
\end{itemize}

An example entry is here:
\begin{keeptogether}
\begin{lstlisting}[language=yaml]
beamExposure:
  name: Beam exposure
  value: 13.8
  units: E22 POT
  uncertainty: negligible
  comment: |
    SNS beam exposure for the first CEvNS detection on liquid argon in terms of protons on target. It represents 6.12 GWhr of integrated beam power.
\end{lstlisting}
\end{keeptogether}

An example entry for a parameter described by a function is here:

\begin{keeptogether}
\begin{lstlisting}[language=yaml]
larQF:
  name: LAr quenching factor
  parameters:
   - name: a
      value: 0.246
      uncertainty: 0.006
    - name: b
      value: 0.00078
      uncertainty: 0.00009
  comment: |
    Form of QF = a + bT(keVnr) where T is the recoil energy in units of keVnr. This value for the QF was determined based on a linear fit to all available data points from the literature in the range 0-125 keVnr. Further described in arXiv:2003.10630.
\end{lstlisting}
\end{keeptogether}

\section{Systematic Errors}
The systematic error PDFs corresponding to the systematic error bands from Analysis A shown in Fig.~4 of Ref.~\cite{Akimov:2020} are also a part of this release. These PDFs represent the various fit systematics obtained in Tab. 1 of Ref.~\cite{Akimov:2020}. The PDFs representing the systematic errors are found in the \texttt{Data/SystErrors} directory of the release. They are provided if the need arises to apply these systematics to models within a separate analysis using the data. The data are fit with alternative PDFs to generate the systematic error envelope. Each bin of a 1-D projection contains a systematic error that is the envelope of the alternative fit results. The systematic PDFs are included in the same binned text file format as the central value PDFs. The set of files used to compute each systematic is given in Tab.~\ref{tab:systematicfiles}, with both a $\pm1\sigma$ systematic PDF given where applicable. The difference between the $\pm1\sigma$ PDFs are provided in the filenames with a '-' in the filename representing $-1\sigma$ and a '+' in the filename representing $+1\sigma$ for a given systematic. The labels given to the systematics is the same as in Tab. 1 of Ref.~\cite{Akimov:2020}. \textbf{Note that every entry in Tab.~\ref{tab:systematicfiles} requires the steady-state background file \texttt{bkgpdf.txt} and the delayed BRN file \texttt{delbrnpdf.txt} within a fit using the systematic error PDFs but is not explicitly written in Tab.~\ref{tab:systematicfiles}}.

For those not interested in performing separate systematic fits to the provided data as part of an analysis, the systematic errors are also included in a 1-dimensional binned tab-separated text file. The lines of these files are formatted as: bin center[units], syst. error min as fraction of bin value, syst. error max as fraction of bin value. For the three dimensions, the corresponding files with the appropriate units are: energy[keVee](\texttt{systerrors1denergy.txt}), $F_{90}$[$F_{90}$](\texttt{systerrors1dpsd.txt}), $t_{\mathrm{trig}}$[$\mu$s](\texttt{systerrors1dtime.txt}). 

\begin{table}
\centering
\begin{tabular}[ht!]{l r r}
    \toprule
	Systematic & CEvNS & Prompt BRN\\
	\midrule
    \multirow{2}{*}{\cevns $F_{90}$ E dependence} & \texttt{cevnspdf-1sigF90.txt} & \multirow{2}{*}{\texttt{brnpdf.txt}} \\ & \texttt{cevnspdf+1sigF90.txt} \\ 
    \cevns $t_{\mathrm{trig}}$ mean & \texttt{cevnspdfCEvNSTimingMeanSyst.txt} & \texttt{brnpdf.txt}  \\ 
    \multirow{2}{*}{BRN E dist.} & \multirow{2}{*}{\texttt{cevnspdf.txt}} & \texttt{brnpdf-1sigEnergy.txt} \\ & & \texttt{brnpdf+1sigEnergy.txt} \\
    \multirow{2}{*}{BRN $t_\mathrm{trig}$ mean} & \multirow{2}{*}{\texttt{cevnspdf.txt}} & \texttt{brnpdf-1sigBRNTimingMean.txt} \\ & & \texttt{brnpdf+1sigBRNTimingMean.txt} \\
    BRN $t_\mathrm{trig}$ width & \texttt{cevnspdf.txt} & \texttt{brnpdfBRNTimingWidthSyst.txt} \\
	\bottomrule
	\end{tabular}
	\caption{Details of systematic PDFs corresponding to the error bands in Fig.~4 of Ref.~\cite{Akimov:2020}. For an analysis, the rows of the table correspond to which PDFs to fit to the data to replicate a certain systematic effect.}
    \label{tab:systematicfiles}
\end{table}

\section{Example Code}
\begin{figure*}
\begin{minipage}{0.325\textwidth}
\includegraphics[width=\textwidth,trim={0 1 40 0},clip]{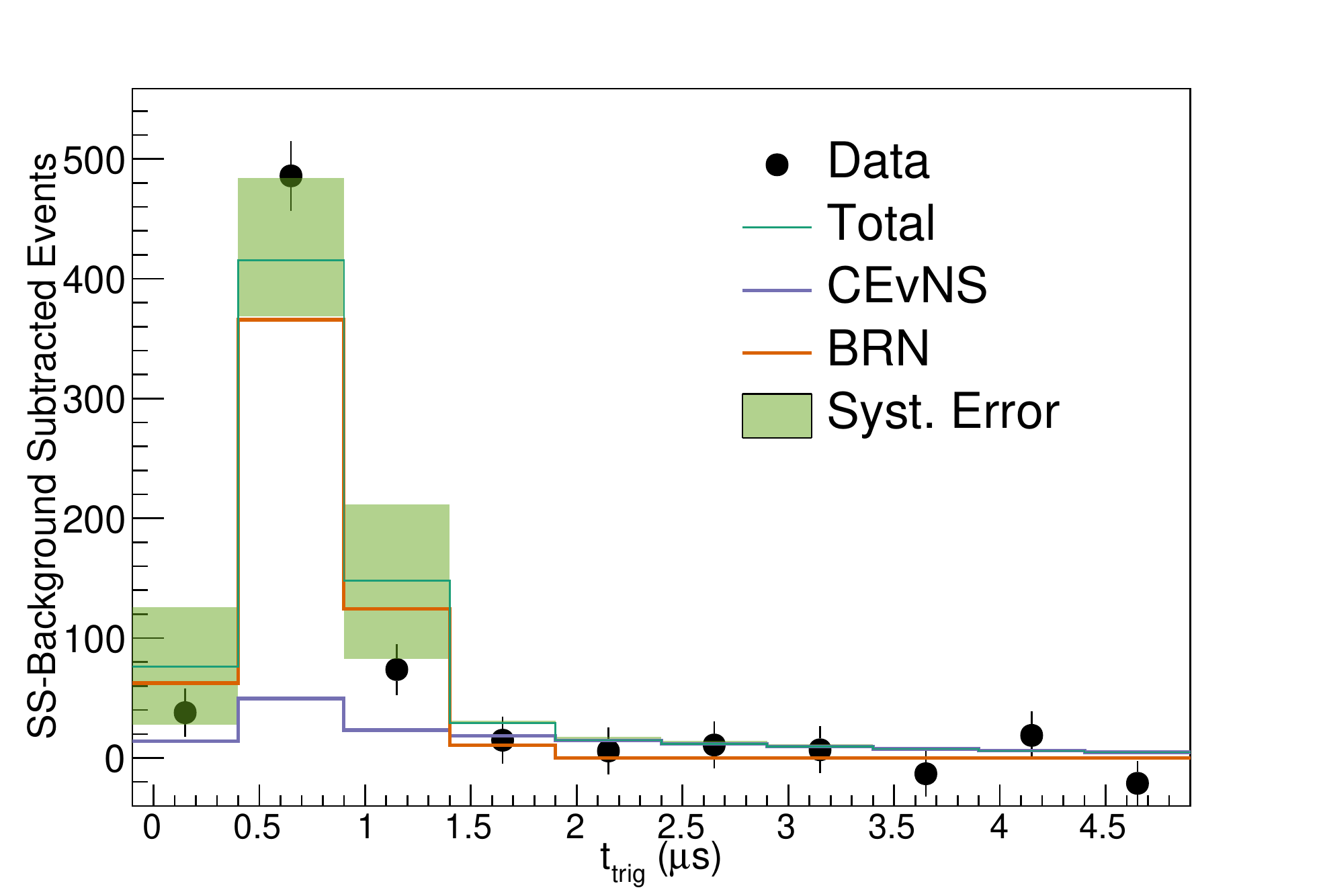}
\end{minipage}
\hfill
\begin{minipage}{0.325\textwidth}
\includegraphics[width=\textwidth,trim={0 1 40 0},clip]{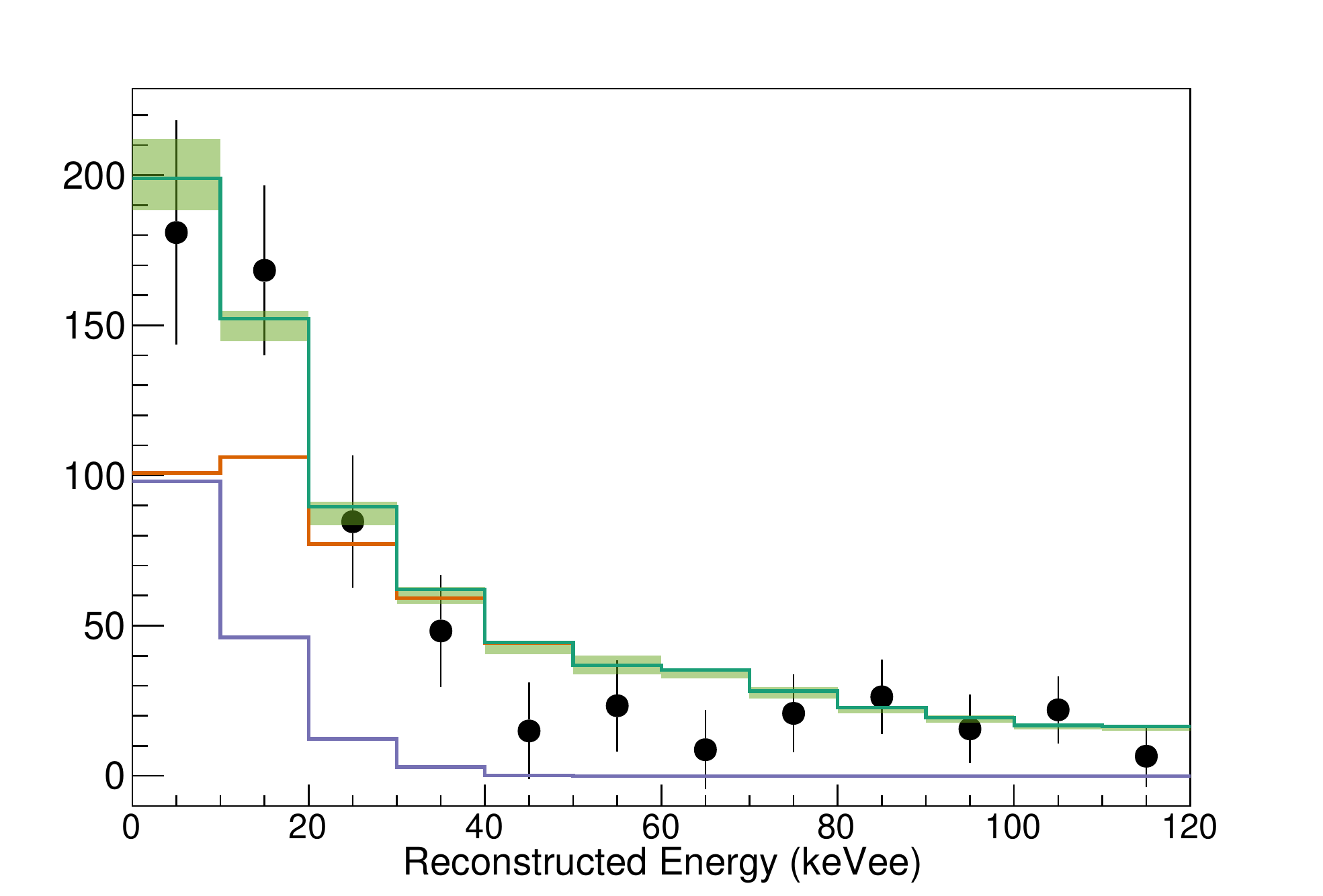}
\end{minipage}
\hfill
\begin{minipage}{0.325\textwidth}
\includegraphics[width=\textwidth,trim={0 1 40 0},clip]{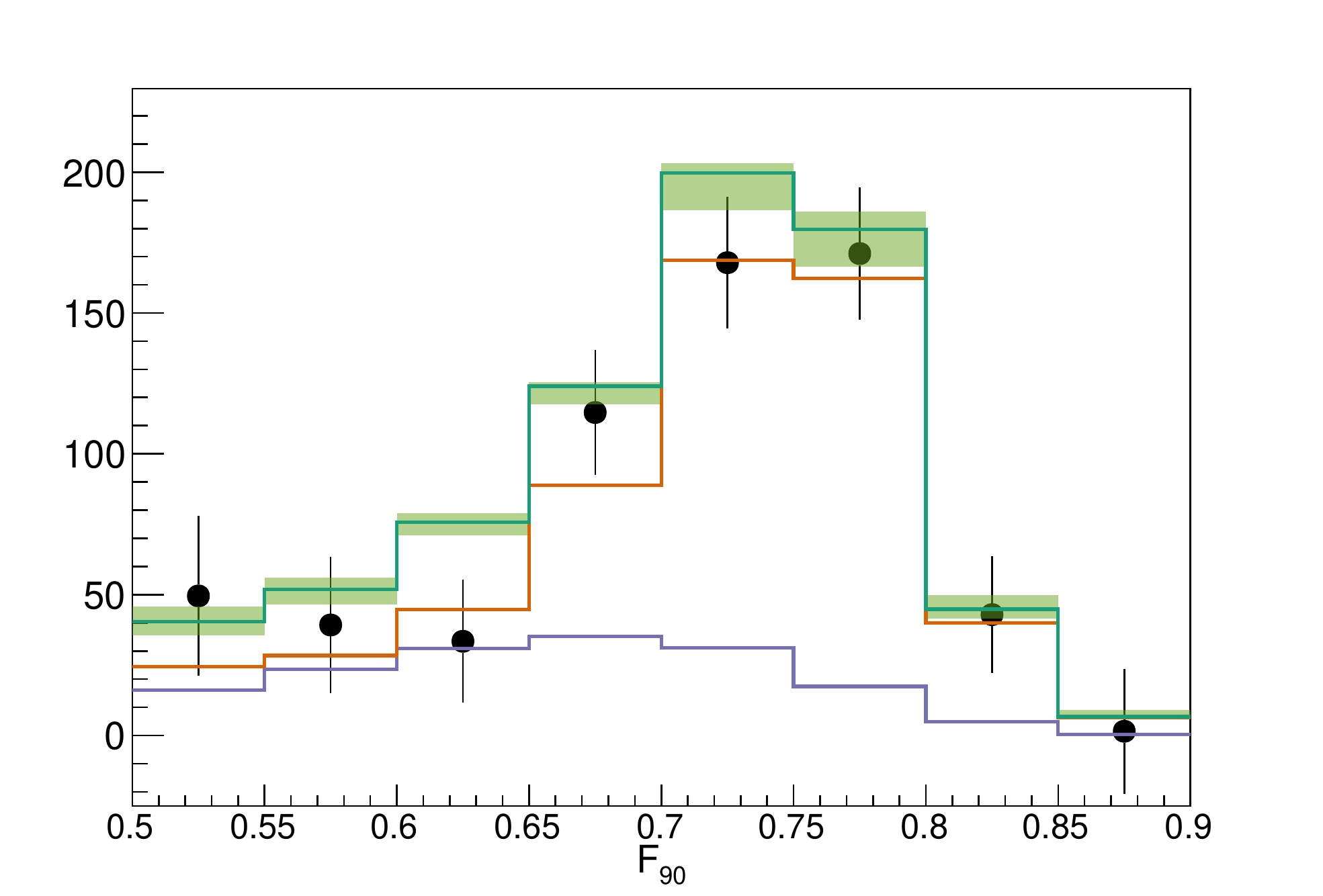}
\end{minipage}
\caption{\label{fig:exprojections} Projection of the best-fit maximum likelihood probability density function (PDF) from Analysis A on $t_\mathrm{trig}$ (left), reconstructed energy (center), and $F_{90}$ (right) along with the data. The example script provided within the data release described in this section creates these projections which replicate Fig.~4 of Ref.~\cite{Akimov:2020}}
\end{figure*}

\texttt{PlotExtractedData.C} is an example ROOT-based macro which parses the text files provided in this release and remakes Fig.~4 of Ref.~\cite{Akimov:2020}. The code is located in the \texttt{ExampleCode} directory of the release. The code gives an example of how to extract the data, compute statistical and systematic errors, and compare the PDFs to the data. The resulting projections generated by this example script are shown in Fig.~\ref{fig:exprojections}.  

\texttt{readYAMLParameters.py} is a Python-based code adapted from the previous CsI data release from the collaboration~\cite{Akimov:2018vzs} which takes in the YAML file \texttt{LArParametersAnlA.yaml} and prints out the name and value of each entry to the command line.

\section{Comments on the use of this release} \label{sec:prescription}

The organization of this release is intended to facilitate an analysis with the same strategy as that described in Ref.~\cite{Akimov:2020}.  An alternative analysis with a different signal hypothesis could proceed with the following steps: 
\begin{enumerate}
\item Prepare alternative signal hypothesis:
\begin{enumerate}
    \item Generate alternative hypothesis event with a given ``true'' nuclear recoil (nr) energy: $E_{nr,true}$(keVnr).
    \item Use included quenching factor to convert to true electron-equivalent (ee) energy: from $E_{nr,true}$(keVnr) to $E_{ee,true}$(keVee).
    \item Use included energy resolution to convert to reconstructed (reco) ee energy: from $E_{ee,true}$(keVee) to $E_{ee,reco}$(keVee).
    \item Create a 1D $E_{ee,reco}$ distribution of the alternative hypothesis (pre-acceptance correction). 
    \item Apply included efficiency curve to produce accepted events $E_{ee,reco}$ distribution (post-efficiency correction).
    \item Then use included \cevns PDFs to determine $F_{90}$ and timing distributions for each $E_{ee,reco}$ bin. The result will be 3D PDF for signal events. 
\end{enumerate}
\item Use provided BRN, steady-state backgrounds 3D PDFs and add to predicted signal. 
\item Use provided binned data with appropriate likelihood procedure to find best fit parameters for the alternative hypothesis and adjusted BRN, SS normalizations.  Note that the overall number of BRN, SS events are not fixed but variable/constrained as reported in Sec.~\ref{sec:included}.
\item Use systematics as reported in Tab.~\ref{tab:systematicfiles} to run a set of alternative fits that allow an extraction of systematic errors on any alternative fit parameters.
\end{enumerate}
Notes:
\begin{itemize}
    \item The PDFs are normalized to the initial CV predictions from Ref.~\cite{Akimov:2020} and a new fit should allow them to vary subject to constraints listed in Tab.~\ref{tab:summaryinfo}. The \cevns PDF was allowed to float.
\end{itemize}

\section{Citing this release} \label{sec:citing}
If you make use of this data release in your work, the COHERENT Collaboration requests that you cite both~\cite{Akimov:2020} in addition to the Zenodo posting of this dataset.
Suggested formatting for the Zenodo citation is given below: 
\begin{displayquote}
	D. Akimov \etal (2020). COHERENT collaboration data release from the first detection of coherent elastic neutrino-nucleus scattering on argon[Data set]. Zenodo. DOI: 10.5281/zenodo.3903810. arXiv: {\tt 2006.12659 [nucl-ex]}.
\end{displayquote}

\bibliographystyle{apsrev}
\bibliography{LArDataRelease}

\end{document}